# Van der Waals interaction of excited media

Yury Sherkunov

Institute for High Energy Density of Joint Institute for High Temperatures, RAS,

Moscow, Russia

Casimir interaction between two media of ground-state atoms is well described with the help of Lifshitz formula depending upon permittivity of media. We will show that this formula is in contradiction with experimental evidence for excited atoms.

We calculate Casimir force between two atoms if one of them or both the atoms are excited. We use methods of quantum electrodynamics specially derived for the problem. It enables us to take into account excited-state radiation widths of atoms. Then we calculate the force between excited atom and medium of ground-state atoms. The results are in agreement with the ones, obtained by other authors who used perturbation theory or linear response theory. Generalization of our results to the case of interaction between two media of excited atoms results in a formula, which is in not only in quantitative, but in qualitative contradiction with the Lifshitz formula. This contradiction disappears if media of ground-state atoms are taken. Moreover, our result does not include permittivity of the media. It includes the quantity which differs from the permittivity only for excited atoms. The main features of our results are as follows. The interaction is resonant; the force may be either attractive or repulsive depending on resonant frequencies of the atoms of different media; the value of the Casimir force may be several orders of magnitude lager than that predicted by the Lifshitz formula. The features mentioned here are in agreement with known experimental and theoretical evidences obtained by many authors for interaction of a single excited atom with dielectric media.

## Introduction

The dispersion force between two atoms separated by a distance $R$ large enough to neglect wave function overlap – the van der Waals or the Casimir force – has been studied in numerous works when the atoms are in ground states. If the distance $R$ is smaller than the wavelength of atom transitions, the force is described by London formula[1]. If $R$ is larger than the wavelength and the retardation effects are significant the force is described by the Casimir formula[2] , which was generalized later to arbitrary distances $R$ by Casimir and Polder[3]. Numerous papers, where the Casimir interaction of ground-state atoms is considered, have been appearing for more than fifty last years. For references see[4 5 6].

If one or both atoms are excited the results for the Casimir force differ significantly from the ones mentioned above. If the atoms are in the ground state the force is attractive. If one of the atoms is excited the force is either attractive or repulsive depending on the transition frequencies of atoms. Moreover, the force is resonant. To obtain these results the authors used either linear response theory[7] , or perturbation methods of quantum electrodynamics[8]. But in both the papers the excited energy level widths of atoms have not been taken into account. But if we deal with excited atoms and resonant interaction, we should take into account the finite level widths of atoms. Here we suggest a method, which enables us to calculate the van der Waals potential taking into account such widths.

Interaction of an excited atom near a cold (non-excited) dielectric surface is of great interest now. There are two theoretical approaches to the problem. The first one is based on linear response theory without explicit quantization of electromagnetic field [9 10]. The second one is based on macroscopic quantum electrodynamics with the permittivity included in the Hamiltonian[11 12]. The review of recent works can be found in the papers [11 12]. Both the approaches result in dependence of the Casimir force on the permittivity of medium. Here the Casimir force is resonant and it can be either attractive or repulsive depending on the relation of excited atom and medium transition frequencies. For dilute gas medium the results are in agreement with the ones obtained for two atoms interaction [10]. The latest experiments [10 13 14] are in agreement with theoretical predictions.

The Casimir force between two dielectric media was found for the first time by Lifshitz [15] with the help of linear response theory. Another method of obtaining Lifshitz's result is based on Matsubara temperature Green functions and is given in the textbook[16]. The results are identical and depend on the permittivities of interacting media. The validity of the Lifshitz formula is discussed now for the case of the interaction between two real metals described by permittivities of the Drude model[17] and two magnetodielectric bodies embedded in another magnetodielectric body[18]. We examine the applicability of the Lifshitz formula to excited media. We will show that the result obtained with the help of the Lifshitz formula for excited media is in contradiction with the results of quantum electrodynamics and, moreover, they are in contradiction with the experimental evidence.

In Section II we consider electric dipole interaction of two atoms one of which is excited. We take into account the radiation width of energy levels. A specially developed method of quantum Green functions is implemented. We show that the results are in agreement with the ones obtained by different authors[1 7 8].

Section III is devoted to interaction of excited atom with dielectric medium of dilute cold gas. We show that the Casimir force is expressed in terms of coherent permittivity but not conventional one. But the results are in agreement with the ones expressed in terms of conventional permittivity[9 10 11 12 13 14]. If a ground-state atom interacts with an excited medium the situation is different. We suppose that such a result cannot be obtained in terms of conventional permittivity.

In Section IV we calculate the Casimir force for a case of two media of diluted gases with excited atoms. The result obtained here is not expressed in terms of conventional permittivity (contrary to the Lifshitz formula) but in terms of coherent permittivity. We have shown that the results obtained with the help of quantum electrodynamics and the Lifshitz formula are not in agreement if the amount of excited atoms is significant. Moreover, the Lifshitz formula is in dramatic contradiction with the theoretical and experimental results obtained for interaction of a single excited atom with cold medium[9 10 11 12 13 14].

## Interaction between an excited atom and a ground-state atom

We consider two nonidentical atoms A and B with infinite masses. We take atom A to be in the excited state and situated at a point with radius-vector $\boldsymbol{R}_A$ and B in the ground state and situated at a point $\boldsymbol{R}_B$. We suppose the electromagnetic field to be in its vacuum state. The exchange interaction is negligible. Let us suppose for the sake of simplicity that the radiation width of excited level of atom A is negligible in comparison with the width of the excited level of atom B. The Hamiltonian of the system is as follows

$$\hat{H} = \hat{H}_A + \hat{H}_B + \hat{H}_{ph} + \hat{H}_{int}, \tag{1}$$

where $\hat{H}_A = \sum_i \varepsilon_{Ai} \hat{b}_i^\dagger \hat{b}_i$, $\hat{H}_B = \sum_i \varepsilon_{Bi} \hat{\beta}_i^\dagger \hat{\beta}_i$ are the Hamiltonians of noninteracting atoms A and B, $\varepsilon_i$ is the energy of i-th state of corresponding atom, $\hat{b}_i(\hat{b}_i^\dagger)$, $\hat{\beta}_i(\hat{\beta}_i^\dagger)$ are annihilation (creation) operators of i-th state of corresponding atom, $\hat{H}_{ph} = \sum_{\boldsymbol{k}\lambda} \omega(\lambda)\left(\hat{\alpha}_{\boldsymbol{k}\lambda}^\dagger \hat{\alpha}_{\boldsymbol{k}\lambda} + \frac{1}{2}\right)$ is the Hamiltonian of free electromagnetic field, $\boldsymbol{k}$ is the wave vector, $\lambda = 1,2,3$ is the index of polarization of electromagnetic field, $\hat{\alpha}_{\boldsymbol{k}\lambda}$ $\left(\hat{\alpha}_{\boldsymbol{k}\lambda}^\dagger\right)$ are annihilation (creation) operators of electromagnetic field,

$$\hat{H}_{int} = -\int \hat{\psi}^\dagger(\boldsymbol{r}-\boldsymbol{R}_A)\hat{d}^\nu \hat{E}^\nu(\boldsymbol{r})\hat{\psi}(\boldsymbol{r}-\boldsymbol{R}_A)d\boldsymbol{r} - \int \hat{\varphi}^\dagger(\boldsymbol{r}-\boldsymbol{R}_B)\hat{d}^\nu \hat{E}^\nu(\boldsymbol{r})\hat{\varphi}(\boldsymbol{r}-\boldsymbol{R}_B)d\boldsymbol{r} \tag{2}$$

is the interaction Hamiltonian, where

$$\hat{\psi} = \sum_i \psi_i\left(\boldsymbol{r} - \boldsymbol{R}_A\right)\hat{b}_i,\ \hat{\varphi} = \sum_i \varphi_i\left(\boldsymbol{r} - \boldsymbol{R}_B\right)\hat{\beta}_i, \tag{3}$$

with $\psi_i\left(\boldsymbol{r} - \boldsymbol{R}_A\right)$ and $\varphi_i\left(\boldsymbol{r} - \boldsymbol{R}_B\right)$ being the wave functions of i-th state of corresponding atoms. $\hat{d}^{\nu}$ is the operator of dipole moment, $\hat{E}^{\nu}\left(\boldsymbol{r}\right)$ is the operator of free electromagnetic field

$$\hat{E}^{\nu}\left(\boldsymbol{r}\right) = i\sum_{\boldsymbol{k}\lambda}\sqrt{\frac{2\pi\omega(\lambda)}{V}}e^{\nu}_{\boldsymbol{k}\lambda}\left(\hat{\alpha}_{\boldsymbol{k}\lambda}e^{i\boldsymbol{k}\boldsymbol{r}} - \hat{\alpha}^{\dagger}_{\boldsymbol{k}\lambda}e^{-i\boldsymbol{k}\boldsymbol{r}}\right), \tag{4}$$

where V is quantization volume, $e^{\nu}_{\boldsymbol{k}\lambda}$ is the polarization unit vector, $\omega(1,2) = k,\ \omega(3) = 0.$

Now our aim is to calculate the van der Waals potential for the system. It is evident that this potential is equal to energy shift of a single atom resulting from the presence of the other atom. Consequently, we should calculate the energy shift of, say, excited atom.

To take into account level widths of atoms we should use a non-perturbative approach. But the methods based upon the linear response theory [9 10] or macroscopic quantum electrodynamics[11,12] are not suitable for us, since these methods involve classical polarizabilities of atoms. In a number of problems these methods yield correct results[9 10 11 12], but, as we are going to show, in general case the van der Waals potential or the Casimir force can not be expressed in terms of classical polarizabilities. To calculate the energy shift we will use method of quantum Green functions similar to the one suggested by L.V. Keldysh for kinetics in a medium [16 19]. This method has no phenomenological elements but, on the other hand, it will be possible to take into account energy level widths of atoms.

Let us consider the excited atom. Let

$$G^{A}_{ll'}\left(x,x'\right) = -i\left\langle \hat{T}_c\hat{\psi}_l\left(x\right)\hat{\psi}^{\dagger}_{l'}\left(x'\right)\hat{S}_c\right\rangle \tag{5}$$

be the Green function of atom A. Here $x = \left\{\boldsymbol{r},t\right\}$, operators are in interaction representation [16],

$$\hat{S}_c = \hat{T}_c\, exp\left\{ \sum_{l=1,2} (-1)^l\, i \int_c \hat{H}_{int\,l}(t)\, dt \right\} \tag{6}$$

is the scattering operator, $c$ is the contour of integration given in fig.1, $\hat{T}_c$ is the operator of time-ordering for contour $c$ [19], $\hat{H}_{int\,l}(t)$ is in interaction representation, $\langle ... \rangle$ means averaging over initial state of free atoms. Using (2), (3) and (4), we obtain

$$\hat{H}_{int\,l}(t) = -\int \hat{\psi}_l^{\dagger}(x)\, \hat{d}^{\nu} \hat{E}_l^{\nu}(x)\, \hat{\psi}_l(x) d\boldsymbol{r} - \int \hat{\varphi}_l^{\dagger}(x)\, \hat{d}^{\nu} \hat{E}_l^{\nu}(x)\, \hat{\varphi}_l(x) d\boldsymbol{r}, \tag{7}$$

where

$$\hat{\psi} = \sum_i \psi_i(\boldsymbol{r} - \boldsymbol{R}_A) e^{-i\varepsilon_{Ai} t} \hat{b}_i\,,\; \hat{\varphi} = \sum_i \varphi_i(\boldsymbol{r} - \boldsymbol{R}_B) e^{-i\varepsilon_{Bi} t} \hat{\beta}_i\,, \tag{8}$$

$$\hat{E}^{\nu}(x) = i \sum_{\boldsymbol{k}\lambda} \sqrt{\frac{2\pi\omega(\lambda)}{V}}\, e^{\nu}_{\boldsymbol{k}\lambda} \left( \hat{\alpha}_{\boldsymbol{k}\lambda} e^{i\boldsymbol{kr}} e^{-i\omega\lambda(\lambda)t} - \hat{\alpha}^{\dagger}_{\boldsymbol{k}\lambda} e^{-i\boldsymbol{kr}} e^{i\omega(\lambda)t} \right). \tag{9}$$

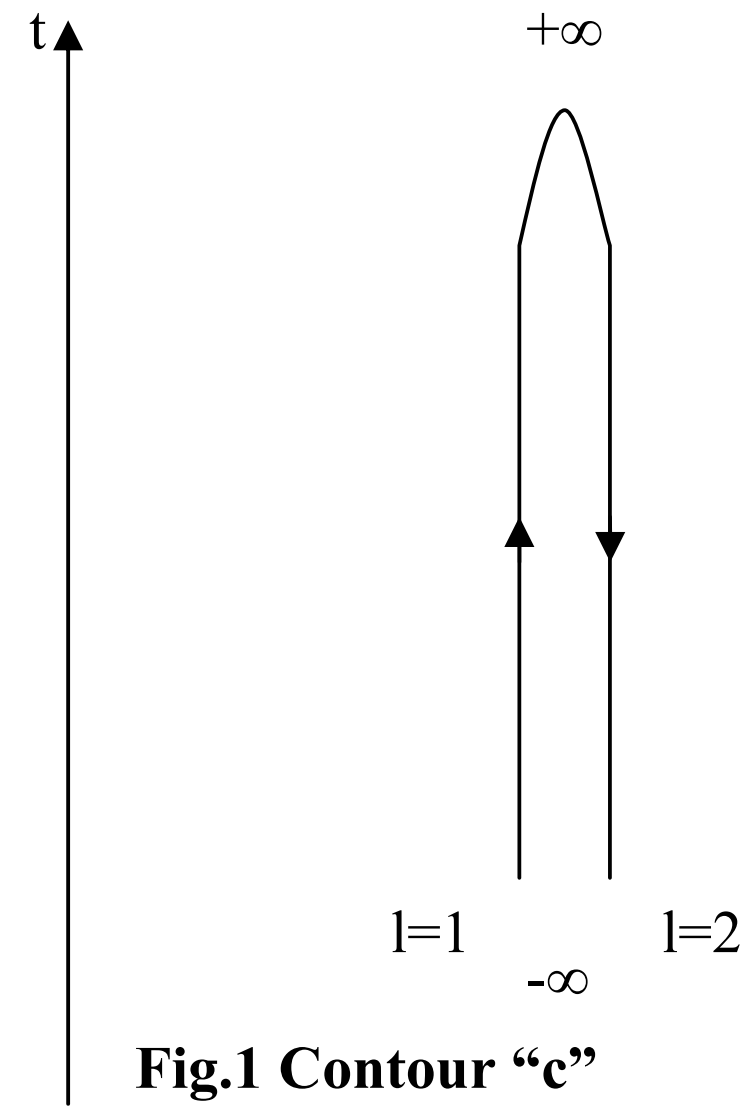

**Fig.1 Contour “c”**

Using the Green function (5) it is easy to find the matrix of density of atom A

$$\rho^A(x, x') = i G_{12}^A(x, x').$$

Representing the S-matrix (6) as a perturbation expansion we come to the following system of equations (Appendix A)

$$\rho^{\mathrm{A}}(x, x') = \rho_c^A(x, x') + \rho_n^A(x, x'). \tag{10}$$

Where $\rho_c^A(x,x')$ represents the coherent channel of interaction, with atom A returning to the initial state (e.g. elastic scattering). Matrix $\rho_n^A(x,x')$ represents the incoherent channel, where atom A does not return to the initial state after interaction (e.g. spontaneous radiation, Raman scattering etc.). Here we are not interested in the incoherent channel processes and we omit $\rho_n^A(x,x')$.

For coherent channel we obtain the equations similar to the ones derived in[20] for electromagnetic field and in[21] for a system of atoms.

$$\begin{aligned}\rho_c^A(x,x') &= \rho_0^A(x,x') + \int dx_1 dx_2 g_{11}^A(x,x_1) M_{11}(x_1,x_2)\rho_0^A(x_2,x') \\ &+ \int dx_1 dx_2 \rho_0^A(x,x_1) M_{22}(x_1,x_2) g_{22}^A(x_2,x') \\ &+ \int dx_1 dx_2 dx_3 dx_4 g_{11}^A(x,x_1) M_{11}(x_1,x_2)\rho_0^A(x_2,x_3) M_{22}(x_3,x_4) g_{22}^A(x_4,x'),\end{aligned} \tag{11}$$

where $\rho_0^A(x,x')$ is the matrix of density of free atom A

$$\rho_0^A(x,x') = \psi_0^*(\boldsymbol{r}')\psi_0(\boldsymbol{r})e^{-i\varepsilon_{A0}(t-t')}, \tag{12}$$

where "0" stands for the initial state of atom A, $g_r^A(x,x')\left(g_a^A(x,x')\right)$ is the retarded (advanced) propagator of atom A, which obeys the following equations[22]

$$\begin{aligned} g_{11}^A(x,x') &= g_{11}^{0A}(x,x') + \int dx_1 dx_2 g_{11}^{0A}(x,x_1) M_{11}(x_1,x_2) g_{11}^A(x_2,x'), \\ g_{22}^A(x,x') &= g_{22}^{0A}(x,x') + \int dx_1 dx_2 g_{22}^{0A}(x,x_1) M_{22}(x_1,x_2) g_{22}^A(x_2,x'). \end{aligned} \tag{13}$$

with

$$\begin{aligned} g_{11}^{0A}(x,x') &= -i\hat{T}_c\left[\hat{\psi}_1(x)\hat{\psi}_1^\dagger(x')\right] = -i\theta(t-t')\sum_i \psi_i^*(\boldsymbol{r}')\psi_i(\boldsymbol{r})e^{-i\varepsilon_{Ai}(t-t')}, \\ g_{22}^{0A}(x,x') &= -i\hat{T}_c\left[\hat{\psi}_2(x)\hat{\psi}_2^\dagger(x')\right] = -i\theta(t'-t)\sum_i \psi_i^*(\boldsymbol{r}')\psi_i(\boldsymbol{r})e^{-i\varepsilon_{Ai}(t-t')} \end{aligned} \tag{14}$$

being the retarded (advanced) propagator of free atom A. $M_{11}$ and $M_{22}$ are the mass operators

$$\begin{aligned} M_{11}(x,x') &= -ig_r^A(x,x')D_{11}^{\nu\nu'}(x',x), \\ M_{22}(x,x') &= -ig_a^A(x,x')D_{22}^{\nu\nu'}(x',x) \end{aligned}, \tag{15}$$

where $D_{11}^{\nu\nu'}(x',x)$ and $D_{22}^{\nu\nu'}(x',x)$ are photon propagators[22]

$$
\begin{aligned}
&D_{11}^{\nu\nu'}(x',x) = D_{11}^{0\nu\nu'}(x',x) + \int dx_1 dx_2 \sum_{\nu_1\nu_2} D_{11}^{0\nu\nu_1}(x',x_1)\Pi_{11}^{\nu_1\nu_2}(x_1,x_2)D_{11}^{\nu\nu_1}(x_2,x), \\
&D_{22}^{\nu\nu'}(x',x) = D_{22}^{0\nu\nu'}(x',x) + \int dx_1 dx_2 \sum_{\nu_1\nu_2} D_{22}^{0\nu\nu_1}(x',x_1)\Pi_{22}^{\nu_1\nu_2}(x_1,x_2)D_{22}^{\nu\nu_1}(x_2,x)
\end{aligned}, \tag{16}
$$

with

$$
\begin{aligned}
&D_{11}^{0\nu\nu'}(x',x) = i\hat{T}_c\left[\hat{E}_1^{\nu}(x')\hat{E}_1^{\nu'}(x)\right], \\
&D_{22}^{0\nu\nu'}(x',x) = i\hat{T}_c\left[\hat{E}_2^{\nu}(x')\hat{E}_2^{\nu'}(x)\right].
\end{aligned} \tag{17}
$$

In frequency-coordinate domain these functions are equal[22]

$$
\begin{aligned}
&D_{11}^{0\nu\nu'}(\omega,\boldsymbol{r}-\boldsymbol{r}') = \omega^2\left[\delta_{\nu\nu'}\left(1+\frac{i}{|\omega||\boldsymbol{r}-\boldsymbol{r}'|}-\frac{1}{\omega^2|\boldsymbol{r}-\boldsymbol{r}'|^2}\right)\right. \\
&\left.+\frac{(\boldsymbol{r}-\boldsymbol{r}')_{\nu}(\boldsymbol{r}-\boldsymbol{r}')_{\nu'}}{|\boldsymbol{r}-\boldsymbol{r}'|^2}\left(\frac{3}{\omega^2|\boldsymbol{r}-\boldsymbol{r}'|^2}-\frac{3i}{|\omega||\boldsymbol{r}-\boldsymbol{r}'|}-1\right)\right]\frac{e^{i|\omega||\boldsymbol{r}-\boldsymbol{r}'|}}{|\boldsymbol{r}-\boldsymbol{r}'|}, \\
&D_{22}^{0\nu\nu'}(\omega,\boldsymbol{r}-\boldsymbol{r}') = \left(D_{11}^{0\nu\nu'}(\omega,\boldsymbol{r}-\boldsymbol{r}')\right)^*
\end{aligned} \tag{18}
$$

Now it is convenient to rewrite the integral equation (11) as a differential one (Appendix B).

$$
\begin{aligned}
&\rho_c^A(x,x') = \Psi(x)\Psi^*(x'), \\
&\left(i\frac{\partial}{\partial t}-\hat{H}_A\right)\Psi(x) = \int M_{11}(x,x_1)\Psi(x_1)\,dx_1.
\end{aligned} \tag{19}
$$

The coherent channel processes do not change the initial state of atom A, consequently

$$
\Psi(x) = \psi_0(\boldsymbol{r}-\boldsymbol{R}_A)f(t), \tag{20}
$$

where index "0" stands for initial state of atom A. Substituting (20) into (19) and neglecting non-diagonal elements of the mass operator we arrive at the following equation

$$
i\frac{\partial}{\partial t}f(t) - \varepsilon_{0A}f(t) = \int_{t_0}^{\infty} M_{11}^{00}(t,t_1)f(t_1)\,dt_1,
$$

$$
M_{11}^{00}(t,t_1) = \int \psi_0^*(\boldsymbol{r}-\boldsymbol{R}_A)M_{11}(x,x_1)\psi_0(\boldsymbol{r}-\boldsymbol{R}_A)\,d\boldsymbol{r}, \tag{21}
$$

here we suppose that the interaction was switched on at $t_0$ ($t_0 \to -\infty$).

Using pole approximation we find

$$\Psi(x)=\psi_0(\boldsymbol{r}-\boldsymbol{R}_A)e^{-i\varepsilon_{A0}t}e^{-iM_{11}^{00}(\varepsilon_{A0})(t-t_0)},$$

where $M_{11}^{00}(\varepsilon_{A0})=\int\limits_{-\infty}^{\infty}M_{11}^{00}(t,t')e^{i\varepsilon_{A0}(t-t')}d(t-t')$ is the Fourier transform of mass operator taken at point $E=\varepsilon_{A0}$.

Thus, the density matrix of coherent channel in energy domain

$$\rho_c^A(E,E',\boldsymbol{r},\boldsymbol{r}')=\int\limits_{t_0}^{\infty}\rho_c^A(x,x')e^{iEt-iE't'}dtdt',\, t_0\to-\infty$$

is

$$\rho_c^A(E,E',\boldsymbol{r},\boldsymbol{r}')=\frac{\psi_0(\boldsymbol{r}-\boldsymbol{R}_A)\psi_0^*(\boldsymbol{r}'-\boldsymbol{R}_A)e^{i(E-E')t_0}}{\left(E-\varepsilon_{A0}-M_{11}^{00}(\varepsilon_{A0})\right)\left(E'-\varepsilon_{A0}-M_{22}^{00}(\varepsilon_{A0})\right)}. \tag{22}$$

Such an equation for the case of electromagnetic field was obtained in[20].

Now we can easily calculate the energy shift of atom A and, consequently, the van der Waals potential

$$U(\boldsymbol{R}_A-\boldsymbol{R}_B)=\Delta E_A=Re\left[M_{11}^{00}(\varepsilon_{A0})\right] \tag{23}$$

and energy level width for atom A resulting from interaction with the vacuum and atom B.

$$\frac{\Gamma_A}{2}=-Im\left[M_{11}^{00}(\varepsilon_{A0})\right]. \tag{24}$$

We suppose that the Lamb shift due to interaction with the vacuum is already taken into account in $\varepsilon_{A0}$ and in the expression (23) we take into account only interaction between atoms A and B.

Using equations(13), (15), and (16) we can draw Feynman's diagrams given in Fig.2.

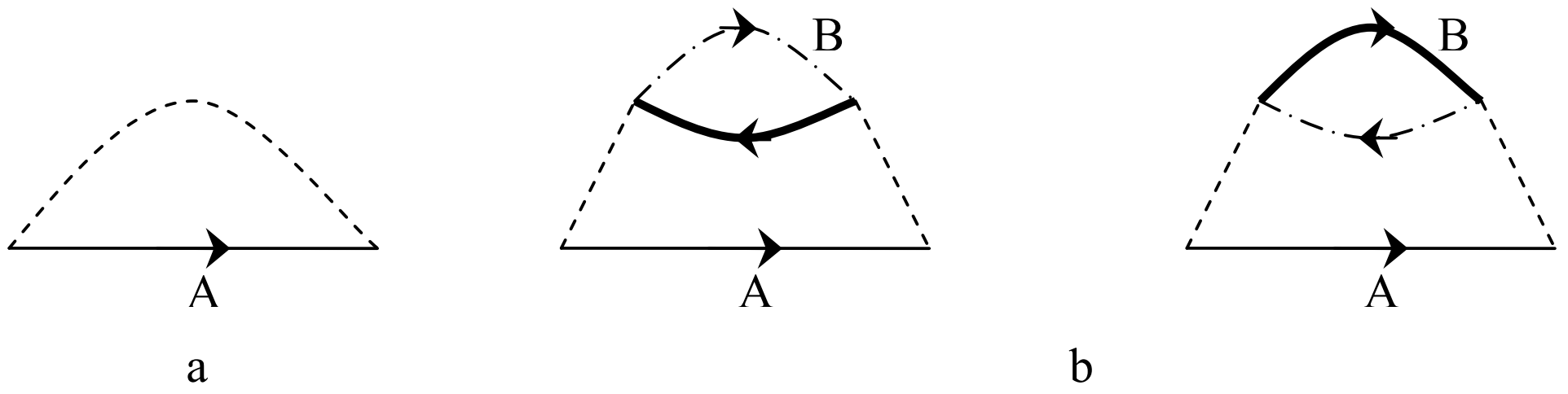

**Fig.2**

Here the solid line corresponds to $g^A$, the dashed line corresponds to $D_{ll'}^{0\nu\nu'}$, the dash-dotted line represents $\rho^B$, the thick solid line represents $g^B$, which are density matrix and propagator of atom B.

We suppose that the ground energy levels of atoms have no width, thus we can replace $\rho^B$ by $\rho_0^B$ and $g^A$ by $g_0^A$. The propagator $g^B$ obeys equation(13). For the sake of simplicity, in equation (13) we take into account only interaction of atom B with the vacuum, which is described by the mass operator given in Fig.2a, where the solid line represents $g_0^B$.

The solution of equation (13) in energy domain is

$$g_{11}^B\left(E,\boldsymbol{r},\boldsymbol{r}'\right)=\sum_i \frac{\varphi_i\left(\boldsymbol{r}-\boldsymbol{R}_B\right)\varphi_i^*\left(\boldsymbol{r}'-\boldsymbol{R}_B\right)}{E-\varepsilon_{Bi}+i\dfrac{\gamma_{Bi}}{2}}, \tag{25}$$

with $\dfrac{\gamma_{iB}}{2}=-Im\left[M_{11}^{ii}\left(\varepsilon_{\mathrm{B0}}\right)\right]$ being the radiation width of energy level *i*, while $M_{11}^{ii}\left(\varepsilon_{\mathrm{B0}}\right)$ is described by the diagram shown in Fig.2.a.

Thus, for the mass operator given in Fig.2 with omitting terms whose contribution to the final result is zero we find

$$
\begin{aligned}
&M_{11}(x,x') = -i\hat{d}^{\mathrm{v}}\hat{d}^{\mathrm{v}'}g_r^{0A}(x,x')D_{11}^{0\mathrm{vv}'}(x',x) \\
&+ig_r^{0A}(x,x')\int \hat{d}^{\mathrm{v}}\hat{d}^{\mathrm{v}'}\hat{d}^{\mathrm{v}_1}\hat{d}^{\mathrm{v}_2}\left(D_{11}^{0\mathrm{vv}_1}(x,x_1)\rho_0^{\mathrm{B}}(x_1,x_2)g_r^{B}(x_2,x_1)D_{11}^{0\mathrm{v}_2\mathrm{v}'}(x_2,x')\right. \\
&\left.+D_{11}^{0\mathrm{vv}_1}(x,x_1)\rho_0^{\mathrm{B}}(x_2,x_1)g_r^{B}(x_1,x_2)D_{11}^{0\mathrm{v}_2\mathrm{v}'}(x_2,x')\right)dx_1dx_2.
\end{aligned} \tag{26}
$$

The first term corresponds to the interaction of atom A with the vacuum (Fig.2a), it results in radiation level width and Lamb shift. Consequently we can omit this term. The second term (Fig.2b) corresponds to the interaction between atoms A and B. In energy domain we have

$$
\begin{aligned}
&M_{11}(E,\boldsymbol{r},\boldsymbol{r}') = \frac{i}{(2\pi)^8}g_r^{0A}(E-\omega,\boldsymbol{r},\boldsymbol{r}') \\
&\times\int \hat{d}^{\mathrm{v}}\hat{d}^{\mathrm{v}'}\hat{d}^{\mathrm{v}_1}\hat{d}^{\mathrm{v}_2}\left(D_{11}^{0\mathrm{vv}_1}(\omega,\boldsymbol{k}_1)\rho_0^{\mathrm{B}}(E'+\omega,\boldsymbol{r}_1,\boldsymbol{r}_2)g_r^{B}(E',\boldsymbol{r}_2,\boldsymbol{r}_1)D_{11}^{0\mathrm{v}_2\mathrm{v}'}(\omega,\boldsymbol{k}_2)\right. \\
&\left.+D_{11}^{0\mathrm{vv}_1}(\omega,\boldsymbol{k}_1)\rho_0^{\mathrm{B}}(E'-\omega,\boldsymbol{r}_2,\boldsymbol{r}_1)g_r^{B}(E',\boldsymbol{r}_1,\boldsymbol{r}_2)D_{11}^{0\mathrm{v}_2\mathrm{v}'}(\omega,\boldsymbol{k}_2)\right)e^{i\boldsymbol{k}_1(\boldsymbol{r}-\boldsymbol{r}_1)}e^{i\boldsymbol{k}_2(\boldsymbol{r}_2-\boldsymbol{r}')}dE'\,d\omega d\boldsymbol{r}_1d\boldsymbol{r}_2.
\end{aligned} \tag{27}
$$

The Fourier transforms of $g_r^{0A}$ and $\rho_0^{\mathrm{B}}$ could be easily found using (12) and (14).

$$
\rho_0^{B}(E,\boldsymbol{r},\boldsymbol{r}') = 2\pi\varphi_0^*(\boldsymbol{r}')\varphi_0(\boldsymbol{r})\delta(E-\varepsilon_{B0}), \tag{28}
$$

$$
g_r^{0A}(E,\boldsymbol{r},\boldsymbol{r}') = \sum_i \frac{\psi_i^*(\boldsymbol{r}')\psi_i(\boldsymbol{r})}{E-\varepsilon_{Ai}+i0}. \tag{29}
$$

Substituting equations (25), (28), and (29) into (27) we find

$$
\begin{aligned}
&M_{11}(E,\boldsymbol{r},\boldsymbol{r}') = \frac{i}{(2\pi)^7}\frac{\psi_g^*(\boldsymbol{r}'-\boldsymbol{R}_A)\psi_g(\boldsymbol{r}-\boldsymbol{R}_A)}{E-\omega-\varepsilon_{Ag}+i0} \\
&\times\int \hat{d}^{\mathrm{v}}\hat{d}^{\mathrm{v}'}\hat{d}^{\mathrm{v}_1}\hat{d}^{\mathrm{v}_2}\left(D_{11}^{0\mathrm{vv}_1}(\omega,\boldsymbol{k}_1)\varphi_g^*(\boldsymbol{r}_2-\boldsymbol{R}_B)\varphi_g(\boldsymbol{r}_1-\boldsymbol{R}_B)\frac{\varphi_e(\boldsymbol{r}_2-\boldsymbol{R}_B)\varphi_e^*(\boldsymbol{r}_1-\boldsymbol{R}_B)}{\varepsilon_{\mathrm{B}g}-\varepsilon_{Be}-\omega+i\frac{\gamma_B}{2}}D_{11}^{0\mathrm{v}_2\mathrm{v}'}(\omega,\boldsymbol{k}_2)\right. \\
&\left.+D_{11}^{0\mathrm{vv}_1}(\omega,\boldsymbol{k}_1)\varphi_g^*(\boldsymbol{r}_1-\boldsymbol{R}_B)\varphi_g(\boldsymbol{r}_2-\boldsymbol{R}_B)\frac{\varphi_e(\boldsymbol{r}_1-\boldsymbol{R}_B)\varphi_e^*(\boldsymbol{r}_2-\boldsymbol{R}_B)}{\varepsilon_{\mathrm{B}g}-\varepsilon_{Be}+\omega+i\frac{\gamma_B}{2}}D_{11}^{0\mathrm{v}_2\mathrm{v}'}(\omega,\boldsymbol{k}_2)\right) \\
&\times e^{i\boldsymbol{k}_1(\boldsymbol{r}-\boldsymbol{r}_1)}e^{i\boldsymbol{k}_2(\boldsymbol{r}_2-\boldsymbol{r}')}d\omega d\boldsymbol{r}_1d\boldsymbol{r}_2d\boldsymbol{k}_1d\boldsymbol{k}_2.
\end{aligned} \tag{30}
$$

Here "g" and "e" stand for ground and excited state correspondingly.

Now we should substitute (30) into (21) and take into account the integral in dipole approximation

$$\int e^{i\boldsymbol{kr}}\psi_i^*\left(\boldsymbol{r}-\boldsymbol{R}\right)\hat{d}^{\nu}\psi_j\left(\boldsymbol{r}-\boldsymbol{R}\right)d\boldsymbol{r}=d_{ij}^{\nu}e^{i\boldsymbol{kR}},$$

where $d_{ij}^{\nu}$ is the matrix element of dipole moment.

$$M_{11}^{00}\left(\varepsilon_{Ae}\right)=\frac{i}{\left(2\pi\right)}\frac{1}{\varepsilon_{Ae}-\omega-\varepsilon_{Ag}+i0}$$
$$\times\int\left(D_{11}^{0\nu\nu_1}\left(\omega,\boldsymbol{R}_B-\boldsymbol{R}_A\right)\frac{d_{eg}^{\nu A}d_{ge}^{\nu' A}d_{eg}^{\nu_1 B}d_{ge}^{\nu_2 B}}{\varepsilon_{Bg}-\varepsilon_{Be}-\omega+i\frac{\gamma_B}{2}}D_{11}^{0\nu_2\nu'}\left(\omega,\boldsymbol{R}_A-\boldsymbol{R}_B\right)\right. \tag{31}$$
$$\left.+D_{11}^{0\nu\nu_1}\left(\omega,\boldsymbol{R}_B-\boldsymbol{R}_A\right)\frac{d_{eg}^{\nu A}d_{ge}^{\nu' A}d_{ge}^{\nu_1 B}d_{eg}^{\nu_2 B}}{\varepsilon_{Bg}-\varepsilon_{Be}+\omega+i\frac{\gamma_B}{2}}D_{11}^{0\nu_2\nu'}\left(\omega,\boldsymbol{R}_A-\boldsymbol{R}_B\right)\right)d\omega.$$

Using the symmetry property of $D_{11}^{0\nu\nu'}$ function $D_{11}^{0\nu\nu'}\left(\omega\right)=D_{11}^{0\nu\nu'}\left(-\omega\right)$, which is evident from (18) we can rewrite the equation (31) in terms of coherent polarizabilities introduces in[22] and widely discussed in [23].

$$M_{11}^{00}\left(\varepsilon_{Ae}\right)=\frac{i}{4\pi}\int_{-\infty}^{\infty}D_{11}^{0\nu\nu_1}\left(\omega,\boldsymbol{R}_B-\boldsymbol{R}_A\right)D_{11}^{0\nu_2\nu'}\left(\omega,\boldsymbol{R}_A-\boldsymbol{R}_B\right)\alpha_A^{(c)\nu\nu'}\left(\omega\right)\alpha_B^{(c)\nu_1\nu_2}\left(\omega\right)d\omega, \tag{32}$$

or using (23) and (24) we find

$$U\left(\boldsymbol{R}_A-\boldsymbol{R}_B\right)=Re\left[\frac{i}{4\pi}\int_{-\infty}^{\infty}D_{11}^{0\nu\nu_1}\left(\omega,\boldsymbol{R}_B-\boldsymbol{R}_A\right)D_{11}^{0\nu_2\nu'}\left(\omega,\boldsymbol{R}_A-\boldsymbol{R}_B\right)\alpha_A^{(c)\nu\nu'}\left(\omega\right)\alpha_B^{(c)\nu_1\nu_2}\left(\omega\right)d\omega\right] \tag{33}$$

and

$$\frac{\Gamma_A}{2}=-Im\left[\frac{i}{4\pi}\int_{-\infty}^{\infty}D_{11}^{0\nu\nu_1}\left(\omega,\boldsymbol{R}_B-\boldsymbol{R}_A\right)D_{11}^{0\nu_2\nu'}\left(\omega,\boldsymbol{R}_A-\boldsymbol{R}_B\right)\alpha_A^{(c)\nu\nu'}\left(\omega\right)\alpha_B^{(c)\nu_1\nu_2}\left(\omega\right)d\omega\right] \tag{34}$$

with the coherent polarizability for the ground state atom

$$\alpha_g^{(c)\nu\nu'}\left(\omega\right)=\frac{d_{ge}^{\nu}d_{eg}^{\nu'}}{\omega_{eg}-\omega-i\frac{\gamma}{2}}+\frac{d_{eg}^{\nu}d_{ge}^{\nu'}}{\omega_{eg}+\omega-i\frac{\gamma}{2}}, \tag{35}$$

for excited atom

$$\alpha_e^{(c)\nu\nu'}(\omega) = \frac{d_{eg}^{\nu} d_{ge}^{\nu}}{-\omega_{eg} - \omega - i\frac{\gamma}{2}} + \frac{d_{ge}^{\nu} d_{eg}^{\nu'}}{-\omega_{eg} + \omega - i\frac{\gamma}{2}}. \tag{36}$$

Here we introduce $\omega_A = \varepsilon_{eA} - \varepsilon_{gA}$, $\omega_B = \varepsilon_{eB} - \varepsilon_{gB}$.

The conventional polarizabilities of atoms are well known [22]

$$\alpha_g^{\nu\nu'}(\omega) = \frac{d_{ge}^{\nu} d_{eg}^{\nu'}}{\omega_{eg} - \omega - i\frac{\gamma}{2}} + \frac{d_{eg}^{\nu} d_{ge}^{\nu'}}{\omega_{eg} + \omega + i\frac{\gamma}{2}}, \tag{37}$$

$$\alpha_e^{\nu\nu'}(\omega) = \frac{d_{eg}^{\nu} d_{ge}^{\nu'}}{-\omega_{eg} - \omega - i\frac{\gamma}{2}} + \frac{d_{ge}^{\nu} d_{eg}^{\nu'}}{-\omega_{eg} + \omega + i\frac{\gamma}{2}}. \tag{38}$$

The equation (33) coincides with the result well known from classical textbook[22] for the case of neglecting the radiation width $\left(\frac{\gamma_B}{2} \to 0\right)$ and supposing that atom A is not excited $(\omega_A \to -\omega_A)$.

The equations (35) and (37) as well as (36) and (38) are different due to the sign of the imaginary part of the denominators of the second terms. The signs of the imaginary parts of the denominators of the conventional polarizabilities (37) and (38) are connected with their analytical properties. They should be analytical in the upper part of the complex plane, while the signs of the imaginary parts of coherent polarizabilities (35) and (36) are the result of coherent principal. If we change the corresponding signs in our calculations we will come to violation of causality principle in quantum electrodynamics[22]. As it was shown in[23 24 25] the signs in the denominators of coherent polarizabilities could be changed only due to the presence of incoherent channel, which describes the processes of spontaneous and induced radiation (at any rate the initial state of atoms should be changed). But in our case of the van der Waals interaction the incoherent channel does not contribute to the result.

After averaging over all possible orientations of dipole moments of atoms we can write [22]

$$d_{eg}^{\nu_1} d_{ge}^{\nu_2} \to \frac{\left|d_{eg}\right|^2}{3} \delta_{\nu_1\nu_2}. \tag{39}$$

Let us consider a case of small distance between the atoms $R << \lambda$, where $R$ is the distance between the atoms and $\lambda$ is the wavelength of radiation of atoms. Substituting (18) into (33) and taking into account (39) as well as $R << \lambda$

$$U(R) = Re\left[\frac{3i}{2\pi R^6}\int_{-\infty}^{\infty}\alpha_A^{(c)}(\omega)\alpha_B^{(c)}(\omega)d\omega\right]. \tag{40}$$

After substituting(35), and (36) into(40), we find

$$U_{eg}(R) = \frac{2}{3R^6}\frac{(\omega_A - \omega_B)\left|d_{eg}^A\right|^2\left|d_{eg}^B\right|^2}{(\omega_A - \omega_B)^2 + \left(\frac{\gamma_B}{2}\right)^2}. \tag{41}$$

For the case of $\left(\frac{\gamma_B}{2} \to 0\right)$ we come to the formula obtained in[7 8]. Here we should mention that the van der Waals interaction between excited and ground-state atoms could be either attractive or repulsive depending on the sign of $\omega_A - \omega_B$.

Now let us consider an opposite case. Let atom A be a ground-state one and atom B be an excited one. In this case we should use equation(26), but we should substitute (22) and (14) into (26) but not (28) and (25). If $t << \tau$, where $t$ is the time of interaction and $\tau$ is the lifetime of excited state of atom B, we come to the evident result

$$U_{ge}(R) = \frac{2}{3R^6}\frac{(\omega_B - \omega_A)\left|d_{eg}^A\right|^2\left|d_{eg}^B\right|^2}{(\omega_A - \omega_B)^2 + \left(\frac{\gamma_B}{2}\right)^2}. \tag{42}$$

The results for the case of both the ground-state atoms could be obtained analogously

$$U_{gg}(R) = -\frac{2}{3R^6}\frac{(\omega_A + \omega_B)\left|d_{eg}^A\right|^2\left|d_{eg}^B\right|^2}{(\omega_A + \omega_B)^2 + \left(\frac{\gamma_B}{2}\right)^2}. \tag{43}$$

Evidently the result (43) corresponds to attraction of atoms. This result coincides with London formula [1] if $\left(\frac{\gamma_B}{2} \to 0\right)$.

## Interaction between an atom and a dielectric surface.

For the sake of simplicity we will consider a dielectric semi-infinite body of dilute gas of atoms. Our aim is to compare the results for the van der Waals force obtained with the help of the Lifshitz formula and the one obtained with the help of quantum electrodynamics taking into account pair interactions between atoms.

1. Let us consider an excited atom A near a surface of a gas of ground-state atoms B. Taking into account only pair interactions we can obtain a formula for the interaction potential by integrating the equation (41) with respect to the volume of the medium, with $\gamma_B$ being a collision width of excited energy level of the atoms of the gas.

$$U_1(z_0) = \int dV \frac{2}{3R^6} \frac{(\omega_A - \omega_B)\left|d_{eg}^A\right|^2 \left|d_{eg}^B\right|^2}{(\omega_A - \omega_B)^2 + \left(\frac{\gamma_B}{2}\right)^2} n,$$

where $n$ is the number of density of atoms of the medium.

If the atom is separated by a distance of $z_0$ from the interface, the result of integrating is

$$U_1(z_0) = \frac{\pi}{9{z_0}^3} \frac{(\omega_A - \omega_B)\left|d_{eg}^A\right|^2 \left|d_{eg}^B\right|^2}{(\omega_A - \omega_B)^2 + \left(\frac{\gamma_B}{2}\right)^2} n, \tag{44}$$

2. Let us consider interaction of ground-state atom A and ground gaseous medium of atoms B. Using expression (43) we find

$$U_2(z_0) = -\frac{\pi}{9{z_0}^3}\left|d_{eg}^A\right|^2 \left|d_{eg}^B\right|^2 \frac{(\omega_A + \omega_B)n}{(\omega_A + \omega_B)^2 + \left(\frac{\gamma_B}{2}\right)^2}. \tag{45}$$

The results (44) and (45) are in agreement with the well known experimental and theoretical results[9 10 11 12 13 14].

3. Let us consider interaction of ground-state atom A and excited gaseous medium of atoms B.

Using expressions (42) and (43) we find

$$U_3(z_0) = \frac{\pi}{9{z_0}^3}\left|d_{eg}^A\right|^2\left|d_{eg}^B\right|^2\left(\frac{(\omega_B-\omega_A)n_e}{(\omega_A-\omega_B)^2+\left(\frac{\gamma_B}{2}\right)^2}-\frac{(\omega_A+\omega_B)n_g}{(\omega_A+\omega_B)^2+\left(\frac{\gamma_B}{2}\right)^2}\right), \tag{46}$$

where $n_e$ and $n_g$ are density numbers of excited and ground-state atoms. The result of such kind, as far as we know, is obtained for the first time.

Using (40) we can rewrite the expressions (44)-(46) in the following way

$$U(z_0) = Re\left[\frac{i}{16\pi z_0^3}\int_{-\infty}^{\infty}\alpha_A^{(c)}(\omega)\left(\varepsilon^{(c)}(\omega)-1\right)d\omega\right]. \tag{47}$$

Here we introduce the coherent permittivity

$$\varepsilon^{(c)}(\omega) = 1+4\pi\left(n_e\alpha_e^{(c)}(\omega)+n_g\alpha_g^{(c)}(\omega)\right), \tag{48}$$

where $\alpha_e^{(c)}(\omega)$ and $\alpha_g^{(c)}(\omega)$ are the coherent polarizabilities given by (35) and (36).

The conventional permittivity could be constructed of the conventional polarizabilities (37) and (38) as follows

$$\varepsilon(\omega) = 1+4\pi\left(n_e\alpha_e(\omega)+n_g\alpha_g(\omega)\right). \tag{49}$$

We see that the van der Waals potential of an atom interacting with a dielectric interface is expressed in terms of coherent permittivity but not in terms of conventional one. But for the case of a single excited atom interacting with a non-excited medium the result obtained by the linear response theory is in agreement with the result obtained by quantum electrodynamics(44). The situation differs dramatically if of a ground-state atom interacts with a medium of excited atoms. The result (46) obtained with the help of quantum electrodynamics can not be obtained in the framework of linear response theory or other phenomenological approaches requiring conventional permittivities to describe media [9 10 11 12]. It results from the dependence of conventional permittivity of media, which must be included in the result obtained with the help of phenomenological approach, on a difference of numbers of density of ground state and excited atoms

$$\varepsilon(\omega)-1=4\pi\left(n_e\alpha_e(\omega)+n_g\alpha_g(\omega)\right)\propto n_g-n_e\,.$$

But the result obtained here without phenomenology (46) does not depend on such a difference.

We must stress that a similar situation, where the result is expressed in terms of coherent permittivity but not conventional one appears in other phenomena. It has been shown[20, 24] that the reflection coefficient of resonant radiation reflected from a gas medium containing excited atoms is expressed in terms of coherent permittivity. Correlation function $\left\langle E^{\nu}\left(\boldsymbol{r}',t\right)E^{\nu'}\left(\boldsymbol{r},t\right)\right\rangle$ of electromagnetic field in a hot medium depends on coherent permittivity as well[23].

## Interaction between two media of excited atoms

Let us consider the simplest case of two media of dilute gases separated by a distance of $L$ (fig.3). Let both the media contain excited atoms.

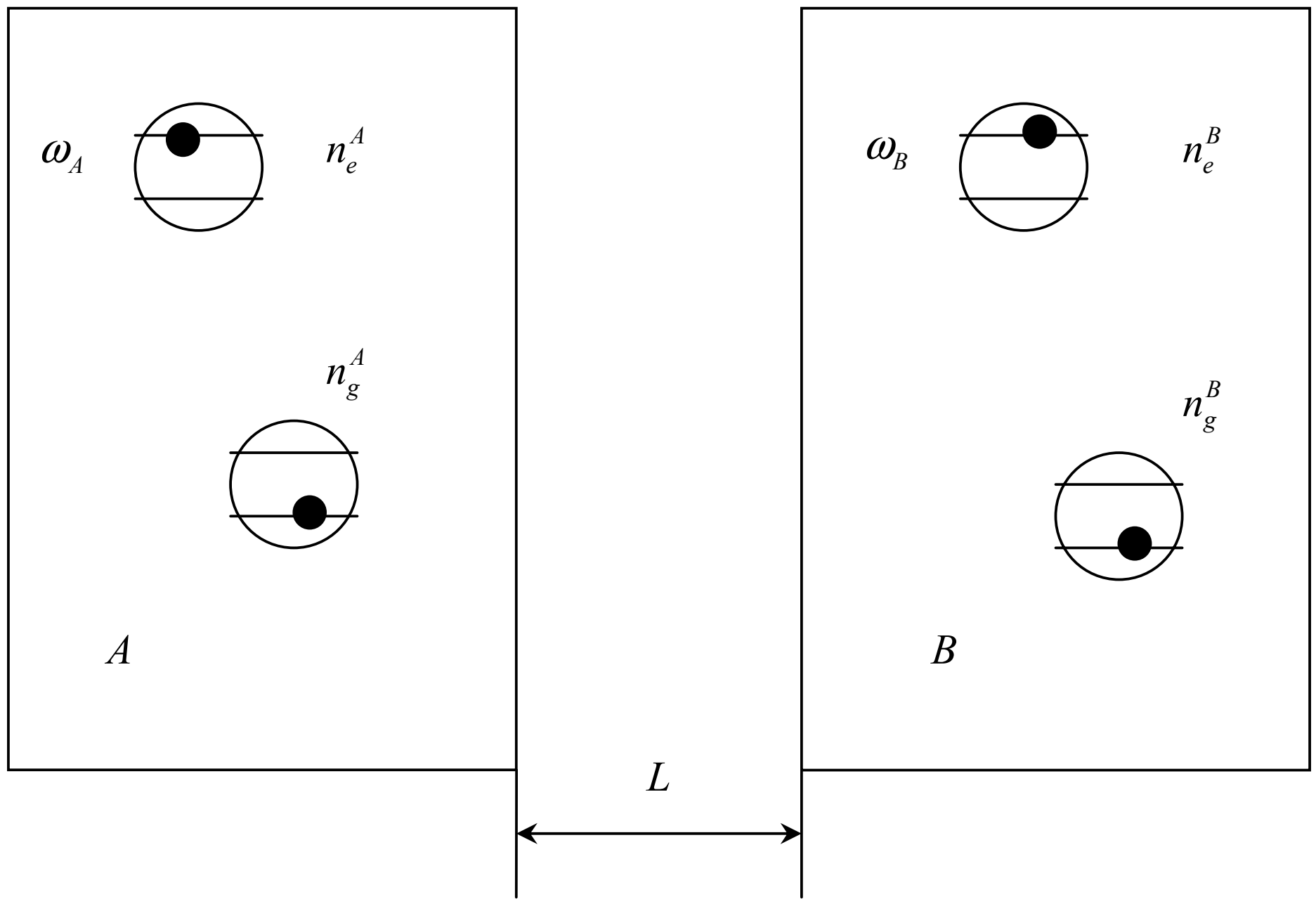

**Fig.3. Interacting media**

To find the van der Waals potential per unit area we should integrate equation (47) with respect to $dz_0$ taking into account pair interactions of atoms of both the media. The result is evident

$$u(L) = Re\left[\frac{i}{128\pi^2 L^2}\int_{-\infty}^{\infty}\left(\varepsilon_A^{(c)}(\omega)-1\right)\left(\varepsilon_B^{(c)}(\omega)-1\right)d\omega\right], \tag{50}$$

where $\varepsilon_A^{(c)}(\omega)$ and $\varepsilon_B^{(c)}(\omega)$ are the coherent permittivities of media *A* and *B*, which are expressed through the coherent permittivities (48). Differentiating (50) by *L* we can find the van der Waals force per unit area

$$F(L) = \frac{\partial}{\partial L}u(L).$$

$$F(L) = Re\left[-\frac{i}{64\pi^2 L^3}\int_{-\infty}^{\infty}\left(\varepsilon_A^{(c)}(\omega)-1\right)\left(\varepsilon_B^{(c)}(\omega)-1\right)d\omega\right]. \tag{51}$$

Substituting(48), (35), and (36) into (51) and calculating the integral with respect to $d\omega$ we find

$$F(L)=\frac{\pi}{9L^3}\left|d_{eg}^A\right|^2\left|d_{eg}^B\right|^2\left(n_g^A n_g^B\frac{(\omega_A+\omega_B)}{(\omega_A+\omega_B)^2+\left(\frac{\gamma_B}{2}\right)^2}-\left(n_e^A n_g^B-n_g^A n_e^B\right)\frac{(\omega_A-\omega_B)}{(\omega_A-\omega_B)^2+\left(\frac{\gamma_B}{2}\right)^2}\right.$$
$$\left.-n_e^A n_e^B\frac{(\omega_A+\omega_B)}{(\omega_A+\omega_B)^2+\left(\frac{\gamma_B}{2}\right)^2}\right). \tag{52}$$

We consider a case of thermal equilibrium, with atoms obeying Boltzman distribution

$$n_e^A=n_g^A e^{-\frac{\omega_A}{T}},\, n_e^B=n_g^B e^{-\frac{\omega_B}{T}},$$

with $n^A=n_g^A+n_e^A,\, n^B=n_g^B+n_e^B$ being the total numbers of density, which is supposed to be constant.

$$F(L,T)=\frac{\pi}{9L^3}\left|d_{eg}^A\right|^2\left|d_{eg}^B\right|^2\frac{n^A n^B}{\left(1+exp\left(-\frac{\omega_A}{T}\right)\right)\left(1+exp\left(-\frac{\omega_B}{T}\right)\right)}$$
$$\times\left(\frac{(\omega_A+\omega_B)\left(1-exp\left(-\frac{\omega_A}{T}\right)exp\left(-\frac{\omega_B}{T}\right)\right)}{(\omega_A+\omega_B)^2+\left(\frac{\gamma_B}{2}\right)^2}-\frac{\left(exp\left(-\frac{\omega_A}{T}\right)-exp\left(-\frac{\omega_B}{T}\right)\right)(\omega_A-\omega_B)}{(\omega_A-\omega_B)^2+\left(\frac{\gamma_B}{2}\right)^2}\right). \tag{53}$$

It is interesting to compare our result (52) with the one derived from the Lifshitz formula [15 16].
For dilute gases the Lifshitz formula is

$$F_L(L)=\frac{1}{32\pi^2L^3}\int_0^\infty\left(\varepsilon_A(iu)-1\right)\left(\varepsilon_B(iu)-1\right)du, \tag{54}$$

with $\varepsilon_A(iu)$ and $\varepsilon_B(iu)$ being the conventional permittivities (49), which are expressed in terms of conventional polarizabilities (37) and (38). After integrating with respect to $du$ supposing that $\omega_A,\omega_B>>\gamma_B$ we find

$$F_L(L)=\frac{\pi}{9L^3}\left|d_{eg}^A\right|^2\left|d_{eg}^B\right|^2\left(n_g^A-n_e^A\right)\left(n_g^B-n_e^B\right)\frac{(\omega_A+\omega_B)}{(\omega_A+\omega_B)^2+\left(\frac{\gamma_B}{2}\right)^2}. \tag{55}$$

The temperature dependence of the Casimir force resulting from the Lifshitz formula is

$$F_L(L,T)=\frac{\pi}{9L^3}\left|d_{eg}^A\right|^2\left|d_{eg}^B\right|^2\frac{n^A n^B\left(1-exp\left(-\frac{\omega_A}{T}\right)\right)\left(1-exp\left(-\frac{\omega_B}{T}\right)\right)}{\left(1+exp\left(-\frac{\omega_A}{T}\right)\right)\left(1+exp\left(-\frac{\omega_B}{T}\right)\right)}\frac{(\omega_A+\omega_B)}{(\omega_A+\omega_B)^2+\left(\frac{\gamma_B}{2}\right)^2}. \quad (56)$$

The difference between expression resulting from quantum electrodynamics (52) and the one obtained with the help of the Lifshitz formula is dramatic. The results coincide only for the case of cold media where density numbers of excited atoms are negligible (fig.4). If the temperatures are high enough for the media to contain excited atoms the dependences (52) and (55) differ qualitatively (Fig.5). Temperature dependences of the Casimir forces are shown in fig.6. Now let us return to the case of interaction of a single excited atom and the dielectric non-excited media discussed in section III. Using expression (55) of the Casimir force obtained by Lifshitz, one can easily find a corresponding expression for the potential of a single excited atom interacting with a cold medium.

$$U_L(z_0)=\frac{\pi}{9{z_0}^3}\left|d_{eg}^A\right|^2\left|d_{eg}^B\right|^2\frac{(\omega_A+\omega_B)n}{(\omega_A+\omega_B)^2+\left(\frac{\gamma_B}{2}\right)^2}. \quad (57)$$

Disagreement of the results of quantum electrodynamics (44) and the consequence of the Lifshitz formula (57) for a case of excited atom near a cold medium is dramatic (fig.7). We see that the van der Waals potential obtained by means of quantum electrodynamics corresponds to resonant attraction (repulsion) for red (blue) detuned atomic transition frequencies $\omega_A < \omega_B\ (\omega_A > \omega_B)$.

This is a well known result. While the van der Waals potential resulting from the Lifshitz formula corresponds to repulsion for all the atom frequencies. The difference between the results is impressive. At some points it is about three orders of magnitude. But our result (44) coincides with the well known theoretical results [9 10 11 12] and experimental ones[10, 13, 14]. Thus, failure of the Lifshitz formula for a case of excited atoms is clear. If we consider a case of a ground-state atom near an interface of cold dielectric, the results obtained by using the Lifshitz formula (55) and quantum electrodynamics (45) are obviously the same.

a)

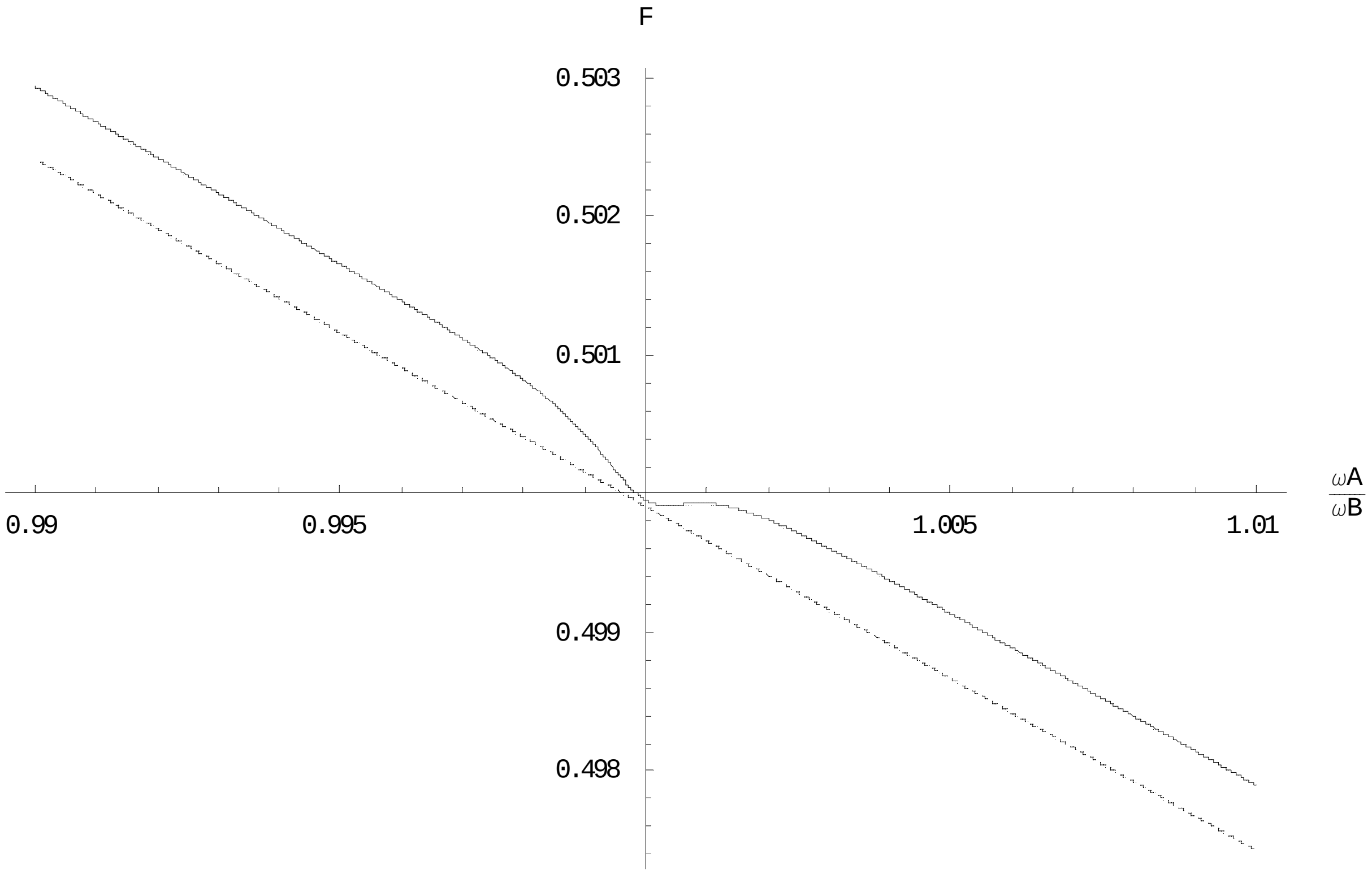

b)

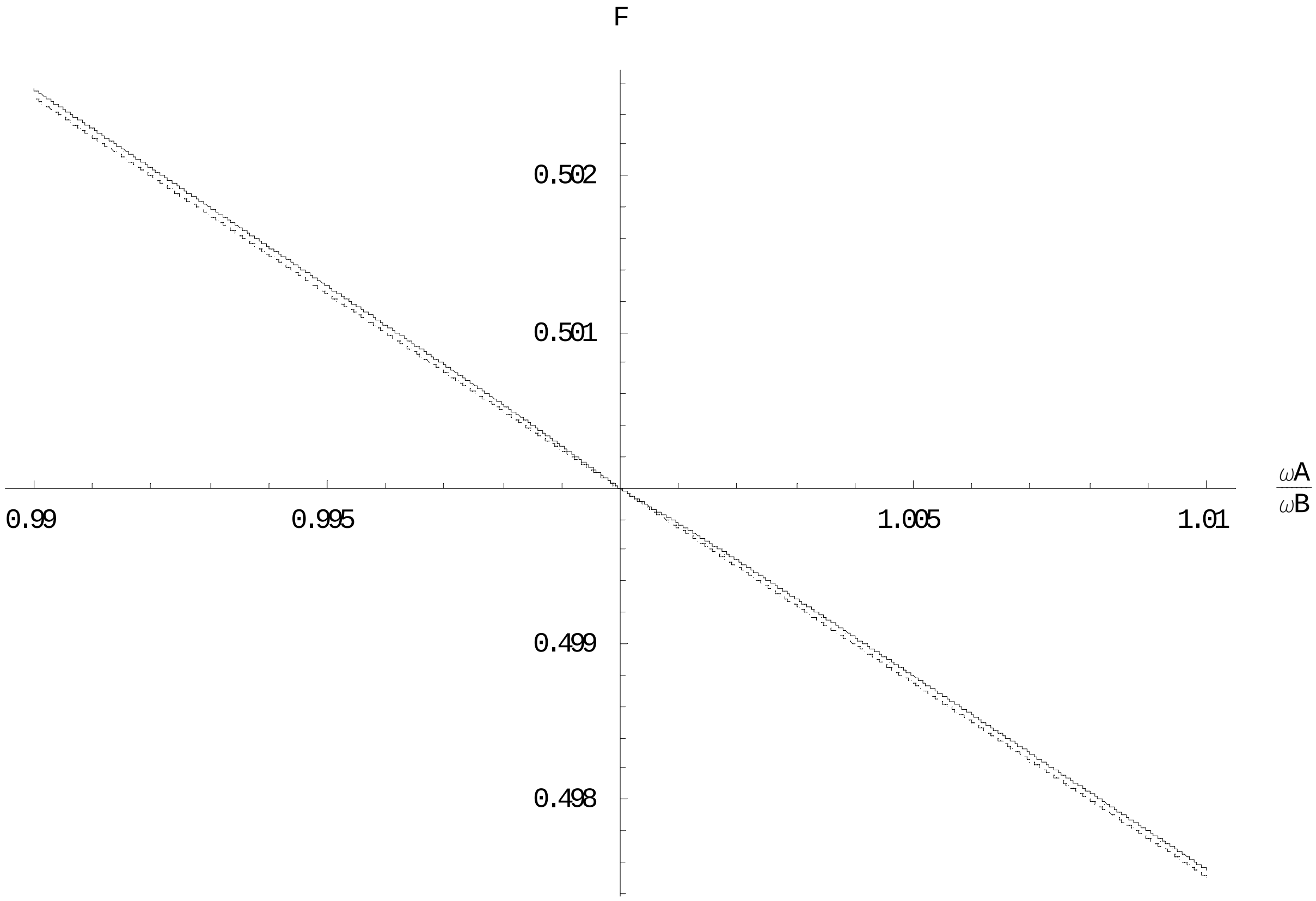

Fig.4

**Normalized Casimir force calculated by means of quantum electrodynamics (53) (solid line) and The Lifshitz formula (56) (dashed line). a)** $T/\omega_B = 0.1, \gamma_B/\omega_B = 0.02$**,**

**b)** $T/\omega_B = 0.08, \gamma_B/\omega_B = 0.02$

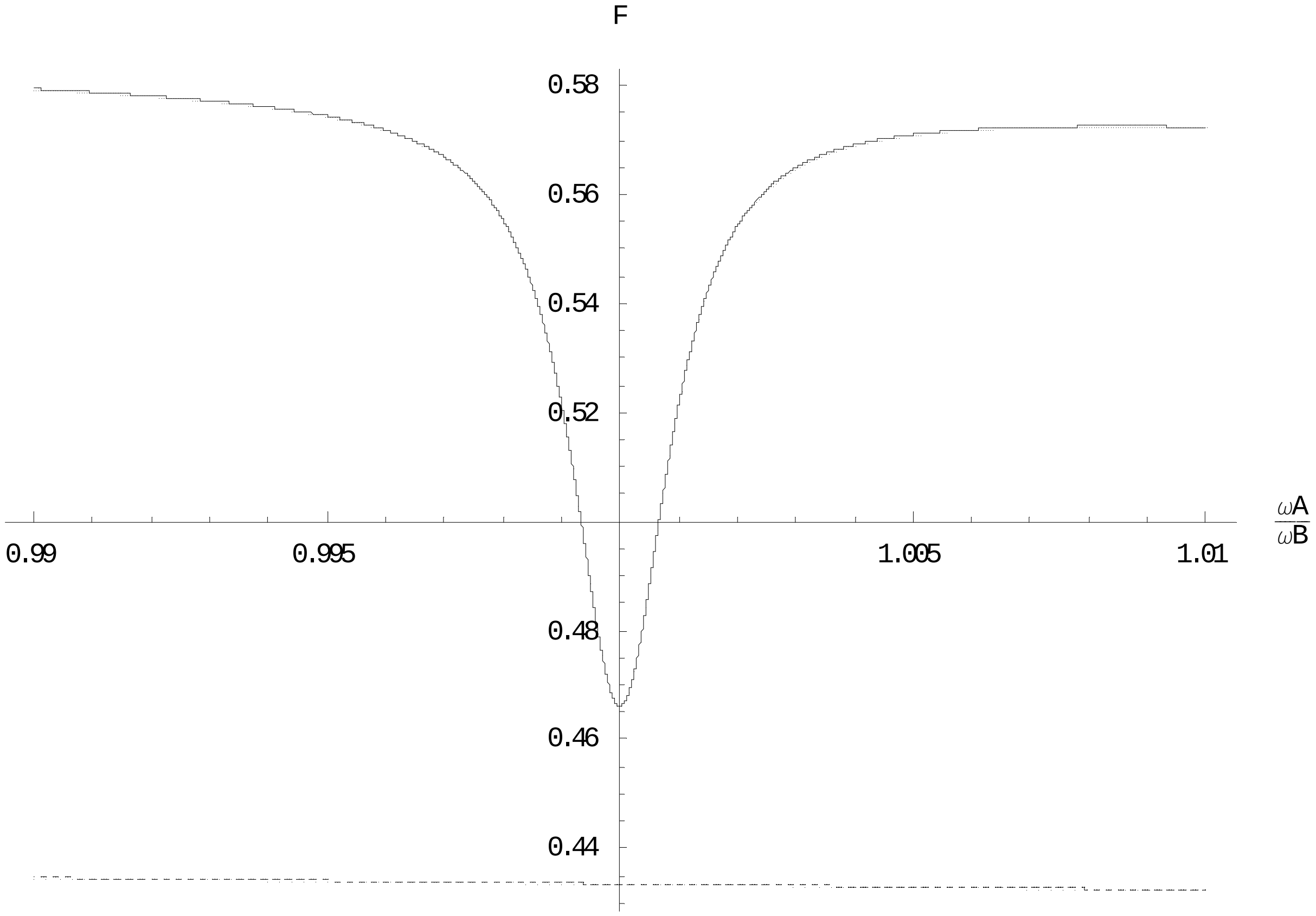

**Fig.5**

**Normalized Casimir force calculated by means of quantum electrodynamics (53)**

**(solid line) and the Lifshitz formula (56) (dashed line).**

**(**$T / \omega_B = 0.3, \gamma_B / \omega_B = 0.02$**)**

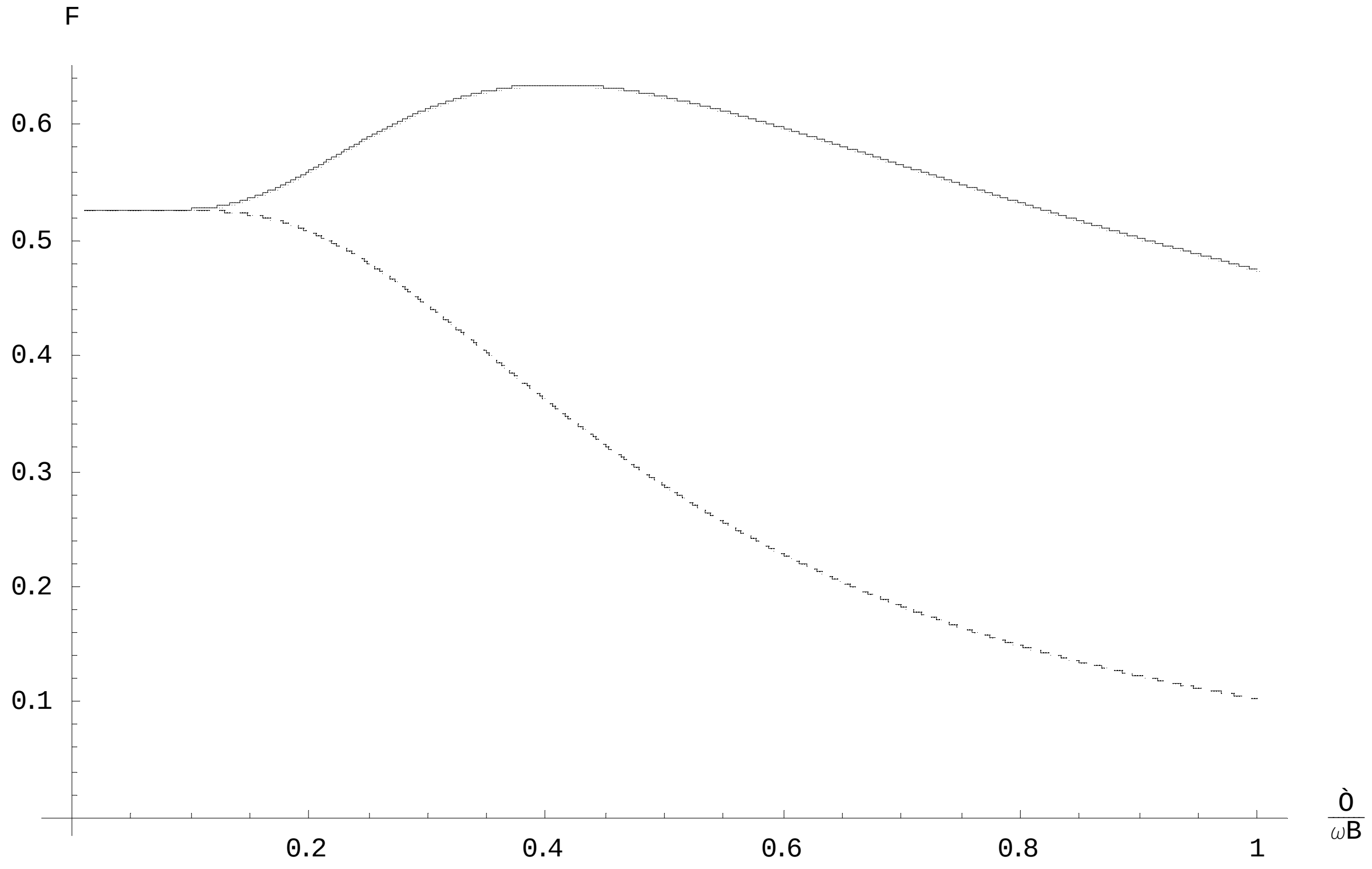

**Fig.6**

**Normalized Casimir force calculated by means of quantum electrodynamics (53) (solid line) and Lifshitz formula (56) (dashed line). (** $\omega_A / \omega_B = 0.9, \gamma_B / \omega_B = 0.02$ **)**

a)

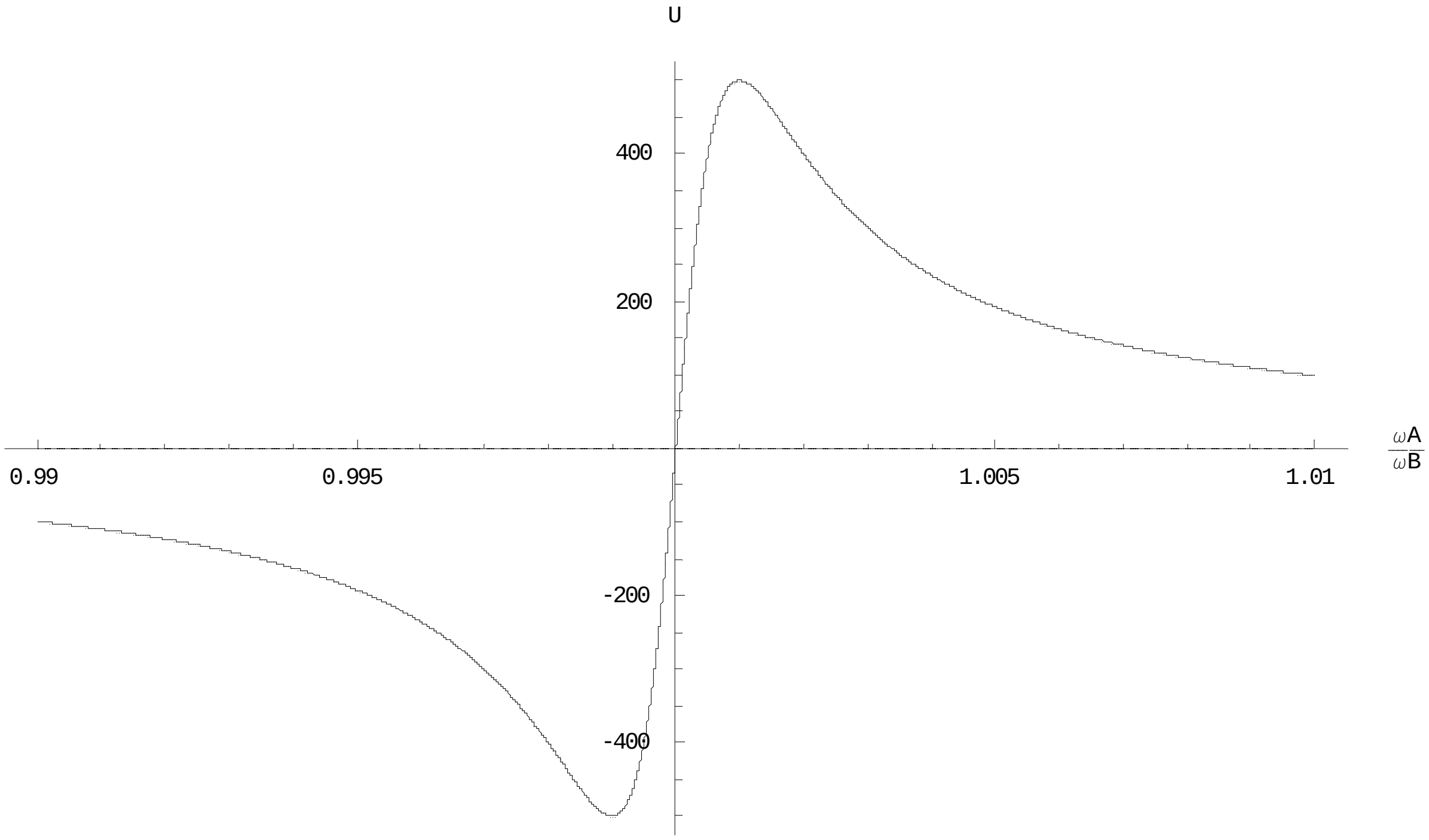

b)

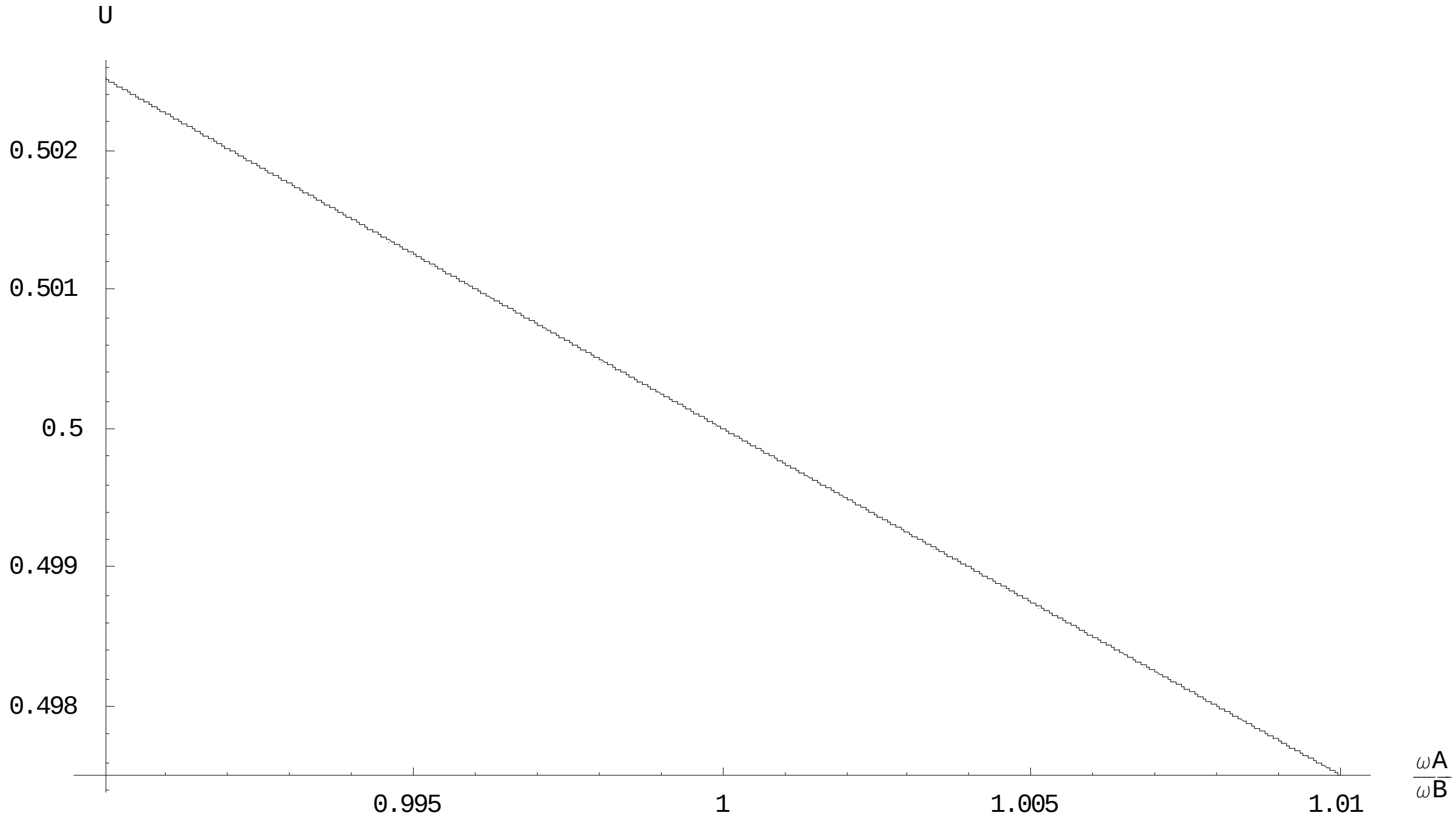

**Fig.7**

**Normalized Casimir potential for excited atom and cold media interaction a) calculated by means of quantum electrodynamics (44) and b) The Lifshitz formula (57). (**$\gamma_B / \omega_B = 0.02$**)**

## SUMMARY

Using a specially developed method of Green functions, which enabled us to take into account energy level widths of atoms, we calculated the van der Waals potential for two atoms dipole-dipole interaction if the atoms are in the following initial states: one atom is excited and the other is in ground state, both the atoms are in ground state. We generalized well known results obtained in the framework of perturbation theory and linear response theory[7 8 22] to the case of finite energy level width of atoms. The results are not expressed in terms of conventional polarizabilities of atoms (37) and (38) but they contain the so-called coherent polarizabilities (35) and (36) with different analytical properties in upper complex semi-plane.

The analysis of the interaction between two atoms enabled us to calculate the van der Waals potential for the interaction of a single atom with a semi-infinite medium. We considered a case of a dilute gas medium and took into account only pair interactions of atoms. The result obtained for the case of excited atom and medium of ground-state atoms is in complete agreement with theoretical works, which used linear response approach or macroscopic quantum electrodynamics (i.e. conventional polarizability)[9 10 11 12], and experimental works[10 13 14], while in our paper it is not expressed in terms of conventional polarizabilities. What is the reason of such an agreement? The authors of the above mentioned papers used the linear response theory or macroscopic QED to describe a medium; as a result, it was described in terms of conventional permittivity(49), while the excited atom was described with the help of Heisenberg equation of motion. Thus, the function corresponding to the excited atom possesses analytical properties of coherent polarizability(36). But for ground-state atoms the first term of conventional polarizability (37) is resonant and it coincides with the first term of coherent polarizability. As a result the formulae obtained in this paper and[9 10 11 12] are in complete agreement.

The situation is different if a ground-state atom is placed in the vicinity of a medium of excited atoms. The result obtained in this paper can not, as far as we know, be obtained with the help of the linear response theory and cannot be expressed in terms of conventional permittivity.

In the last chapter we compared the results obtained with the help of the Lifshitz formula and quantum electrodynamics for the van der Waals interaction of two media of excited atoms. Here we considered dilute gas media and took into account only pair interactions. It was shown that the results coincide only if the media do not contain excited atoms. If the concentrations of excited atoms are significant, the difference of the results is dramatic. The result obtained in this paper is expressed in terms of coherent permittivity, while the Lifshitz formula depends on conventional permittivity. We compared the results of the Lifshitz formula calculated for a case of a single excited atom near ground-state medium and showed that it is not in agreement with theoretical and experimental results[9 10 11 12 13 14]. The difference is dramatic. Quantum electrodynamics results in *resonant* interaction (attractive or repulsive), while the Lifshitz formula gives us *non-resonant* repulsion only. The graphs are given in Fig.7. The difference may be up to three orders of magnitude.

Thus, we state that the Lifshitz formula is applicable only for ground-state (cold) media. If the media are hot enough to possess excited atoms in significant amount, the Casimir interaction can not be described by the Lifshitz formula, at any rate, for the distances smaller than the wave length of atom transition. It can not be described even in terms of conventional permittivities of media. To describe the Casimir interaction of excited media one should use coherent permittivities.

## Appendix A. Method of Green functions.

Here we use a specially elaborated method of quantum Green functions, which enables us to take into account energy level width of atoms. This method resembles a very well known method of kinetic Green's functions suggested by L.V.Keldysh[19].

Now we will outline some basic principles of diagram technique.

A Green function of atom A is given by equation (5), with (6) being scattering operator taken on the contour given in fig. 1. Density matrix of atom A is

$$\rho^{A}\left(x,x'\right)=iG_{12}^{A}\left(x,x'\right).$$

To derive equations (10) and (11) we should expand the scattering operator (6) and substitute the result into equation (5). One should mention that all odd orders of the expansion are equal to zero since we have only one operator of electromagnetic field in each term of the interaction Hamiltonian (7). The first two orders of perturbation theory read

$$\rho^A(x,x') = \rho_0^A(x,x')$$
$$-\frac{1}{2}\left\langle \hat{T}_c \int dx_1 dx_2 \hat{\psi}_{1}(x)\hat{\psi}_2^+(x')(-1)^{l_1+l_2}\hat{\psi}_{l_1}^+(x_1)\hat{E}_{l_1}^{\nu_1}(x_1)\hat{d}^{\nu_1}\hat{\psi}(x_1)\hat{\psi}_{l_2}^+(x_2)\hat{E}_{l_2}^{\nu_2}(x_2)\hat{d}^{\nu_2}\hat{\psi}(x_2)\right\rangle.$$

Using Wick's theorem we find

$$\begin{aligned} \rho^A(x,x') &= \rho_0^A(x,x') \\ &-i\int dx_1 dx_2 \rho_0^A(x,x_1)\hat{d}^{\nu}\hat{d}^{\nu'} g_{22}^{0A}(x_1,x_2) D_{22}^{0\nu\nu'}(x_2,x_1) g_{22}^{0A}(x_2,x') \\ &-i\int dx_1 dx_2 g_{11}^{0A}(x,x_1)\hat{d}^{\nu}\hat{d}^{\nu'} g_{11}^{0A}(x_1,x_2) D_{11}^{0\nu\nu'}(x_2,x_1) \rho_0^A(x_2,x') \\ &-i\int dx_1 dx_2 g_{11}^{0A}(x,x_1)\hat{d}^{\nu}\hat{d}^{\nu'} \rho_0^A(x_1,x_2) D_{21}^{0\nu\nu'}(x_2,x_1) g_{22}^{0A}(x_2,x'), \end{aligned} \tag{58}$$

with the Green functions given by equations (14) and (17).

We take into account that for a single atom all the normal products of orders higher than two are equal to zero

$$\left\langle \hat{N}\hat{\psi}_{l_1}(x)\hat{\psi}_{l_2}^+(x')\hat{\psi}_{l}^+(x_1)\hat{\psi}_{l'}(x_2)\ldots\right\rangle = 0,$$

while the second order of normal product represent the density matrix of initial state of atom

$$\rho^0(x,x') = \left\langle \hat{N}\hat{\psi}_{l_1}(x)\hat{\psi}_{l_2}^+(x')\right\rangle.$$

We take into account that $g_{12}^0 = -i\left\langle \hat{\psi}^\dagger\hat{\psi}\right\rangle_{vacuum} = 0$.

We can draw Feynman's diagram corresponding to equation ( 58) (fig.8)

a

b

c

d

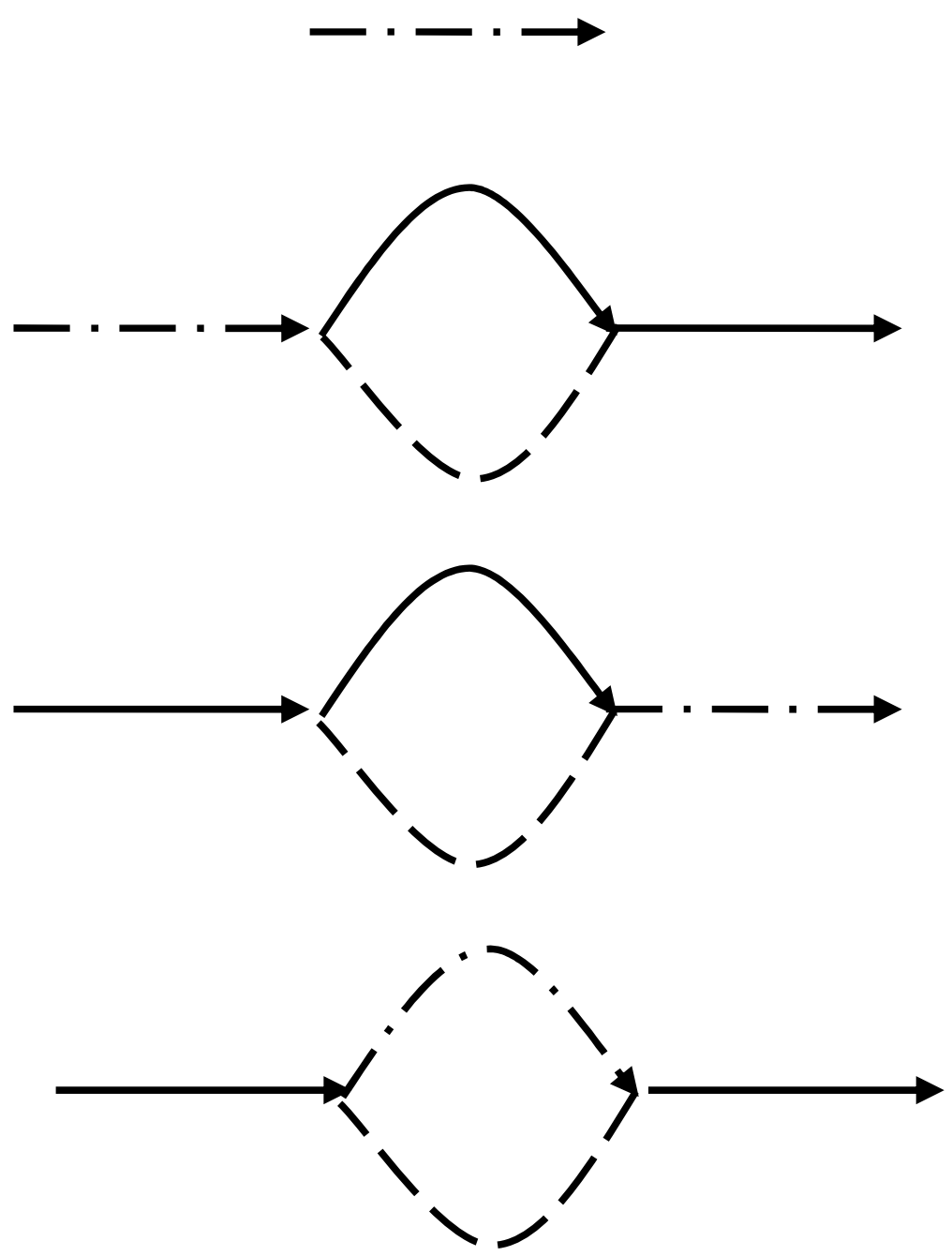

**Fig.8 Feynman's diagram corresponding to equation ( 58). a) first term, b) second term, c) third term, d) fourth term. Solid line corresponds to $g^{0A}$, dashed line corresponds to $D_{ll'}^{0\nu\nu'}$, dash-dotted line represents $\rho_0^A$.**

All the disconnected diagrams are canceling.

The first term of the equation ( 58) is represented by diagram (a), it corresponds to the matrix of density of the initial state of the atom, the second and the third terms are represented as diagrams (b) and (c). They correspond to the processes of coherent channel, with the resultant state of atom being the same as the initial one; it means that the resultant state is described by the same wave function. As an example of such processes we can consider elastic scattering of a photon, or interaction of an atom with the electromagnetic vacuum. The last term given by diagram (d) represents the process of incoherent channel, with the initial state changing as a result of such a process. As an example we may consider the processes of inelastic scattering of a photon, or spontaneous and induced radiation of an atom. It is significant that these channels (coherent and incoherent) are separated and the complete matrix of density is equal to the sum of contributions of these channels. The same separation of channels appears in Γ-operator technique[20].

Now we take into account the higher orders of perturbation technique and use the well known Dyson equation for photon and electron propagators (13) and (16). It is easy to show that we should substitute complete electron and photon propagators satisfying the Dyson equations (13) and (16) into the equation ( 58) instead of free field propagators and add a term appearing in the fourth order of perturbation technique and shown in fig.9. Neglecting the incoherent channel, which has nothing to do with the van der Waals interaction, we come to the equation(11), with the mass operators given by formulae (15).

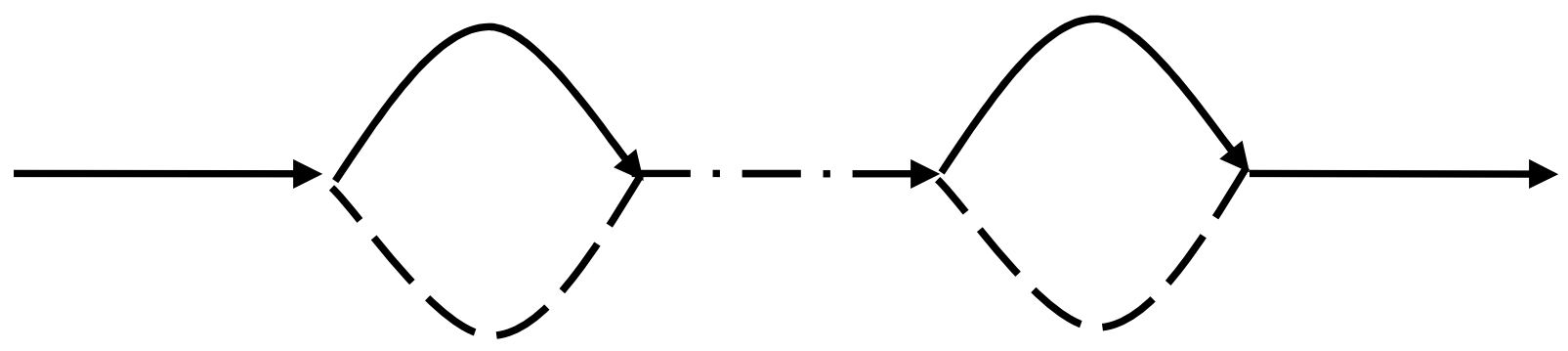

**Fig.9. Feynman's diagram of the fourth order of coherent channel**

## Appendix B Derivation of equation (19)

Here we will derive the differential equation (19) using the integral one (11).

It is easy to show[22] that the free electron propagators and the density matrix satisfy the equations

$$\begin{aligned} g_{11}^{0(-1)} g_{11}^{0}\left(x,x'\right) &= \delta\left(x-x'\right), \\ g_{22}^{0}\left(x,x'\right) g_{22}^{0(-1)} &= \delta\left(x-x'\right), \end{aligned} \qquad (59)$$

$$\begin{aligned} g_{11}^{0(-1)} \rho^{0}\left(x,x'\right) &= 0, \\ \rho^{0}\left(x,x'\right) g_{22}^{0(-1)} &= 0, \end{aligned} \qquad (60)$$

$$g_{11}^{0A,B(-1)} = \left(i\frac{\partial}{\partial t} - \hat{H}_{A,B}\right), \quad g_{22}^{0A,B(-1)} = \left(i\frac{\partial}{\partial t'} - \hat{H}_{A,B}\right) \qquad (61)$$

Using equations ( 59) - ( 61) and (13) we find

$$\begin{aligned} g_{11}^{0A(-1)} g_{11}^{A}(x,x') &= \delta(x-x') + \int dx_2 M_{11}(x,x_2)\, g_{11}^{A}(x_2,x'), \\ g_{22}^{A}(x,x')\, g_{22}^{0A(-1)} &= \delta(x-x') + \int dx_1 g_{22}^{A}(x,x_1)\, M_{22}(x_1,x'). \end{aligned} \quad (62)$$

Now the equation (11) can be rewritten

$$\begin{aligned} & g_{11}^{0A(-1)} \rho_c^{A}(x,x')\, g_{22}^{0A(-1)} = \\ & \int dx_2 dx_3 M_{11}(x,x_2)\, \rho_0^{A}(x_2,x_3)\, M_{22}(x_3,x') \\ & + \int dx_1 dx_2 dx_3 dx_4 M_{11}(x,x_2)\, \rho_0^{A}(x_2,x_3)\, M_{22}(x_3,x_4)\, g_{22}^{A}(x_4,x_1)\, M_{22}(x_1,x') \\ & + \int dx_1 dx_2 dx_3 dx_4 M_{11}(x,x_4)\, g_{11}^{A}(x_4,x_1)\, M_{11}(x_1,x_2)\, \rho_0^{A}(x_2,x_3)\, M_{22}(x_3,x') \\ & + \int dx_1 dx_2 dx_3 dx_4 dx_5 dx_6 M_{11}(x,x_5)\, g_{11}^{A}(x_5,x_1)\, M_{11}(x_1,x_2)\, \rho_0^{A}(x_2,x_3)\, M_{22}(x_3,x_4)\, g_{22}^{A}(x_4,x_6)\, M_{22}(x_6,x'). \end{aligned}$$

Using Dyson equations (13) and formula (11) we come to the following equation

$$g_{11}^{0A(-1)} \rho_c^{A}(x,x')\, g_{22}^{0A(-1)} = \int dx_1 dx_2 M_{11}(x,x_1)\, \rho_c^{A}(x_1,x_2)\, M_{22}(x_2,x'), \quad (63)$$

The equation ( 63) can be easily solved if one represents the density matrix as $\rho_c^{A}(x,x') = \Psi(x)\Psi^{*}(x')$, with $\Psi(x)$ being the wave function of atom A in Shrödinger picture. Such a representation is evident since the coherent channel describes the processes which return the atoms to the initial states, consequently the final state of atoms can be described in terms of wave functions (pure state) if the initial state is pure. Taking into account the formulae ( 61) we come to the equation (19).

## Acknowledgments

This work would not be carried out without permanent attention and valuable advice by Prof. Boris A. Veklenko.

I acknowledge useful discussion at the seminar of the Theoretical department of the Institute of General Physics under supervision of Prof. A.A.Rukhadze.

I thank Dr.S.Y.Buhmann for papers [11 12] and discussion of some aspects of the Casimir interaction.

[1] F. London, Z. Phys. **63**, 245 (1930).

[2] H.B.G. Casimir, Proc. K. Ned. Akad. Wet. Ser. B **51**, 793 (1948).

[3] H. B. G. Casimir and D. Polder, Phys. Rev. **73**, 360 (1948).

[4] Yu. S. Barash and V. L. Ginzburg, Sov. Phys. Usp. **18**, 305 (1975).

[5] G. Plunien, B. Muller, and W. Greiner, Phys. Rep. **134**, 87 (1986).

[6] P. W. Milonni and M.- L. Shih, Phys. Rev. A **45**, 4241 (1992).

[7] E. A. Power and T. Thirunamachandran, Phys. Rev. A **47**, 2539 (1993).

[8] E. A. Power and T. Thirunamachandran, Phys. Rev. A **51**, 3660 (1995).

[9] J. M. Wylie and J. E. Sipe, Phys. Rev. A **30**, 1185 (1984).

[10] M. Fichet, F. Shuller, D. Bloch, and M. Ducloy, Phys. Rev. A **51**, 1553 (1995).

[11] L. Knoll, S. Scheel, and D.-G. Welsch in *Coherence and Statistics of Photons and Atoms*, edited by J. Perina (Wiley, New York, 2001).

[12] S. Y. Buhmann, L. Knoll, D.- G. Welsch, and Ho Trung Dung, Phys. Rev. A **70**, 052117 (2004).

[13] H. Failache, S. Saltiel, M. Fitchet, D. Bloch, and M. Ducloy, Phys. Rev. Lett. **83**, 5467 (1999).

[14] H. Failache, S. Saltiel, A. Fischer, D. Bloch, and M. Ducloy, Phys. Rev. Lett. **88**, 24603-1 (2002).

[15] E. M. Lifshitz, Sov. Phys. JETP **2**, 73 (1956).

[16] E. M. Lifshitz and L.P.Pitaevski, *Statistical physics. Part 2* (Butterworth-Heinenmann, Oxford, 1980).

[17] V. B. Bezerra, G. L. Klimchitskaya, V. M. Mostepanenko, and C. Romero, Phys. Rev. A **69**, 022119-(1-9) (2004).

[18] Ch. Raabe and D.- G. Welsch, Phys. Rev. A **71**, 013841 (2005).

[19] L. V. Keldysh, Sov. Phys. JETP **20**, 1018 (1964).

[20] B. A. Veklenko, Sov. Phys. JETP **69**, 258 (1989).

[21] B. A. Veklenko, Izvestiia Vuzov SSSR. Fizika. (in Russian) **5**, 57 (1987).

[22] V.B. Berestetski, E.M. Lifshitz, and L.P. Pitaevski, *Quantum electrodynamics* (Pergamon, Oxford, 1982).

[23] B. A. Veklenko and Yu. B. Sherkunov, Condensed Matter Physics **7**, 603 (2004).

[24] B. A. Veklenko, R. B. Gusarov, and Yu. B. Sherkunov, JETP **86**, 289 (1998).

[25] B. A. Veklenko and Yu. B. Sherkunov, JETP **89**, 821 (1999).